\documentclass[pre,preprint,floatfix,showpacs]{revtex4}

\usepackage{graphicx}
\usepackage{dcolumn}
\usepackage{bm}

\begin{document}


\title{Detrended fluctuation analysis of power-law-correlated sequences
with random noises}

\author{Shin-ichi Tadaki}
\email{tadaki@cc.saga-u.ac.jp}
\affiliation{%
Computer and Network Center,
Saga University,
Saga 840-8502, JAPAN
}%

\date{\today}

\begin{abstract}
Improvement in time resolution sometimes introduces short-range random
noises into temporal data sequences.  These noises affect the results
of power-spectrum analyses and the Detrended Fluctuation Analysis
(DFA). The DFA is one of useful methods for analyzing long-range
correlations in non-stationary sequences.  The effects of noises are
discussed based on artificial temporal sequences.  Short-range noises
prevent power-spectrum analyses from detecting long-range
correlations.  The DFA can extract long-range correlations from noisy
time sequences.  The DFA also gives the threshold time length, under
which the noises dominate. For practical analyses, coarse-grained time
sequences are shown to recover long-range correlations.
\end{abstract}

\maketitle

\section{Introduction}

Studies of temporal data with long-range correlations recently have
attracted research interests in various fields of physics, biology,
social sciences, technologies and so on.  Researchers have been trying
to observe power-law fluctuations in various temporal data.  And the
origins of those power-law fluctuations have been one of hot research
subjects.  Increment of the amount of such data enables us to
understand complex systems based on data obtained empirically.

Those temporal data observed in complex systems are sometimes not
stationary.  The detrended fluctuation analysis (DFA) is one of the
methods for analyzing non-stationary sequences for detecting
long-range correlations.  It was first developed for analyzing the
long-range correlations in deoxyribonucleic acid (DNA)
sequences\cite{Peng:1994,Peng:1995}.  The method has been employed for
observing their power-law properties in various time series with
non-stationarity

The first step in the DFA method is to define the profile as the
accumulated deviation from the average of the data.  The data sequence
is divided into non-overlapping segments of equal length $l$.  Fitting
the profile by polynomials in each segment defines the local trend.
If the local trend is obtained as a line, the DFA method is called the
first order DFA.  We employ the first order DFA in this paper for
simplicity.

Then we evaluate the standard deviation $F(l)$ of the profile from
the local trend.  If the data sequence has power-law fluctuations,
namely the power spectrum $P(k)$ of the data obeys the power-law
\begin{equation}
P(k)\sim k^{-\gamma},
\end{equation}
the dependence of $F(l)$ on the segment length $l$ is given as
\begin{equation}
F(l)\sim l^{\alpha},\ \gamma = 2\alpha -1.
\end{equation}

Increment of the amount of data sometimes means improvement in time
resolution.  How does the improvement in time resolution contribute to
understand long-range correlations in data sequences?  Let us consider
data traffic in the Internet, for instance.  Internet traffic had been
thought to be uncorrelated and be modeled by a Poisson process,
because hosts are assumed to send data packets randomly.  The validity
of this assumption has clearly lost on the basis of various
experimental measurements\cite{Paxson}.  Power-law properties of
Internet traffic have been investigated
instead\cite{Csabai,Takayasu:1994,Tadaki:2007}.  

In general, Internet traffic data is collected by a software tool
called MRTG (Multi Router Traffic Grapher)\cite{MRTG}.  It
communicates with routers and switches through SNMP (Simple Network
Management Protocol).  With its default setting, the MRTG collects the
amount of packets as 5 minutes average.  By shortening the period for
collecting data, various types of irregularity will be included:
asynchronous behavior of clients and routers, external noises such
as behavior of users, and statistical errors of the observation.  The
Internet traffic, in fact, has been reported to be random in smaller
time scale than 100 ms\cite{Ribeiro2005315}.

The purpose of this paper is to understand how the randomness or
irregularity in short time scales affects results of the DFA and
power-spectrum analyses.  For investigating the effects of short-range
noises, in this work, artificial data with long-range correlation and
short-range randomness are prepared.  The results with the DFA and
standard power spectra on the artificial data will be investigated.

The organization of this paper is as follows: The Fourier Filter
Method (FFM) is employed to generate time sequence with a power-law
correlation in \S 2.  The results of the standard power-spectrum and
the DFA method are investigated.  The short time scale random noises
are introduced into the time sequence in \S 3 by changing the filter
function in FFM.  The power-spectrum analysis will be investigated to
be affected strongly by the noise.  The practical way to eliminate
noises is averaging over short-time scales.  The coarse grained
sequence is investigated in \S 4.  Section 5 is devoted to summary and
discussion.

\section{Fourier Filter Method}
We generate artificial time sequences with power-law correlations for
observing the effects of short-range noises on time sequences with
long-range correlations.  The Fourier Filter Method
(FFM)\cite{Peng:1991,Prakash:1992} is one of the methods for
generating such sequences.  The method is so simple that we can
introduce various types of spectra into sequences.  The method was
improved for extending the range of correlation\cite{Makse:1996}. We
employ the original form of the method for simplicity.

An uncorrelated random sequence of length $T$ is prepared as $\lbrace
u_t\rbrace$ ($t=0,1,\ldots,T-1$) in FFM.  The correlation function of
this sequence is given by
\begin{equation}
C_\tau = \frac{1}{T} \sum_{t=0}^{T-1} u_t u_{t+\tau}.
\end{equation}
The Fourier components of the correlation is given as
\begin{equation}
\hat{C}_k = \frac{1}{\sqrt{T}} 
\sum_{\tau=0}^{T-1} e^{-2\pi i k \tau/T} C_\tau
=\frac{1}{\sqrt{T}} \hat{u}_k \hat{u}_{-k},
\end{equation}
where $\hat{u}_k$ is a Fourier component of the sequence $\lbrace u_t
\rbrace$.  The sequence $\lbrace u_t\rbrace$ is prepared randomly. 
So the Fourier components of the correlation are almost flat.

A correlation will be implemented into the sequence by changing
amplitudes of the Fourier components $\lbrace\hat{u}_k\rbrace$.  To
introduce a power-law correlation, a filter is defined
\begin{equation}
S(k) = k^{-\gamma}.
\end{equation}
The new sequence $\lbrace \eta_t\rbrace$ is defined with
its Fourier component $\hat{\eta}_k$ and the filter.
\begin{equation}
\hat{\eta}_k= S^{1/2}(k) \hat{u}_k
\end{equation}
The new sequence $\lbrace \eta_t\rbrace$ bears power-law fluctuations
\begin{equation}
P(k)\sim k^{-\gamma}.
\end{equation}

\begin{figure}[ht]
\begin{center}
\resizebox{0.45\textwidth}{!}{
\includegraphics{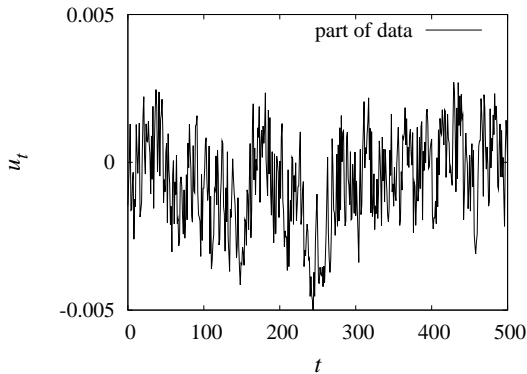}}
\caption{A part of the sequence generated by FFM with $\gamma=0.9$.}
\label{dataWithoutCutoff.eps}
\end{center}
\end{figure}

We generate a sequence with length $T=2^{20}\sim 10^{6}$ in this
paper.  A part of the generated sequence by FFM with $\gamma=0.9$ is
shown in Fig.~\ref{dataWithoutCutoff.eps}.  It does not look like a
simple random sequence.  There seems to be long range correlations.

\begin{figure}[ht]
\begin{center}
\resizebox{0.45\textwidth}{!}{\includegraphics{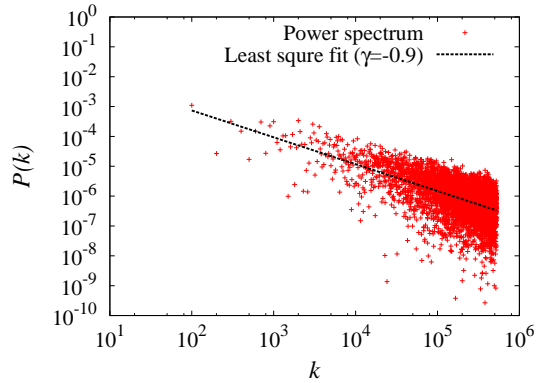}}
\caption{The power spectrum of the sequence generated by FFM
with $\gamma=0.9$.  The line shows the result by the least square
method for fitting all data points, which gives the expected value
$\gamma=0.90$.  Note that the number of data points is reduced.}
\label{fftWithoutCutoff.eps}
\end{center}
\end{figure}

The power-spectrum analysis is one of the most standard methods for
detecting power-law properties in time sequences.  The power spectrum of
the new sequence $\eta_t$ is shown in Fig.~\ref{fftWithoutCutoff.eps}.
It shows clear power-law dependence.  The least square method for
fitting all data points gives the expected value of the power exponent
$\gamma=0.90$.

\begin{figure}[ht]
\begin{center}
\resizebox{0.45\textwidth}{!}{\includegraphics{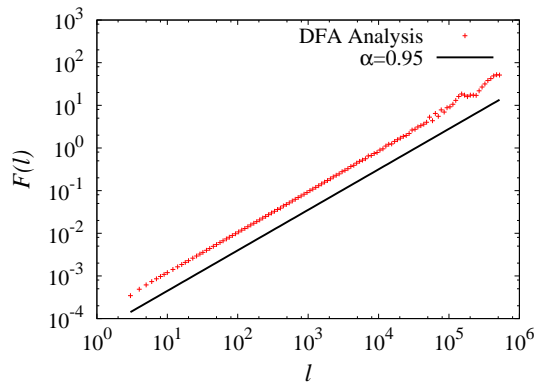}}
\caption{The result of the DFA analysis of the sequence 
generated by FFM with $\gamma=0.9$.}
\label{dfaWithoutCutoff.eps}
\end{center}
\end{figure}

Figure \ref{dfaWithoutCutoff.eps} shows the result of the DFA
analysis.  The observed exponent $\alpha$ corresponds to the expected
value $\alpha=(\gamma+1)/2=0.95$. Namely, if the power-law correlation
covers the whole range of the data, the result of the power spectrum
coincides with the result of the DFA.

\section{Fourier Filter Method with short-range noise}

\begin{figure}[ht]
\begin{center}
\resizebox{0.45\textwidth}{!}{\includegraphics{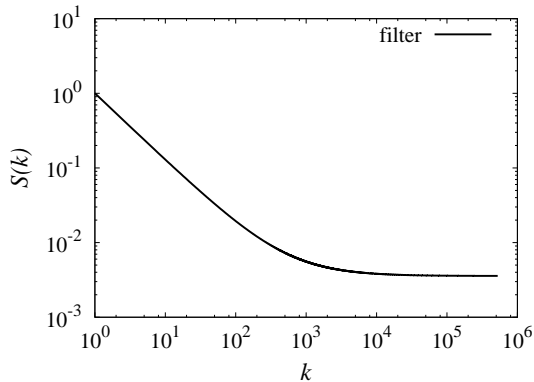}}
\caption{Fourier filter with threshold for $k_c=10^{-3}T$.}
\label{filter.eps}
\end{center}
\end{figure}

Various types of irregularity will be included in data by improving
time resolution of observation.  To investigate the effect of
short-range irregularity on long-range correlations, let us change the
filter $S(k)$ as follows for including short-range noises:
\begin{equation}
S(k)=k^{-\gamma} + k_c^{-\gamma},
\end{equation}
where $k_c$ is a constant corresponding to a threshold of the filter.
The filter $S(k)$ is almost constant for larger wave numbers than
$k_c$.  The randomness in $\lbrace u_k\rbrace$, namely, is not
suppressed in larger wave numbers than $k_c$.  The fluctuation of the
new sequence will obey a power-law in the longer range ($k<k_c$), but
is random in the shorter range ($k>k_c$).  Figure \ref{filter.eps}
shows the filter with threshold $S(k)$ for $k_c = 10^{-3}T$.

\begin{figure}[ht]
\begin{center}
\resizebox{0.45\textwidth}{!}{\includegraphics{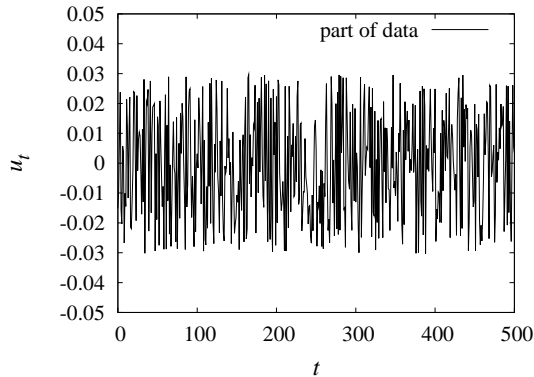}}
\caption{A part of the sequence generated by FFM with threshold.}
\label{data.eps}
\end{center}
\end{figure}

Figure \ref{data.eps} shows a part of the generated sequence by FFM
with threshold.  The amplitudes of high frequency random modes are
larger than those in the sequence by FFM without threshold.  So the
sequence looks random at a glance, by comparing
Fig.~\ref{dataWithoutCutoff.eps}.

\begin{figure}[ht]
\begin{center}
\resizebox{0.45\textwidth}{!}{\includegraphics{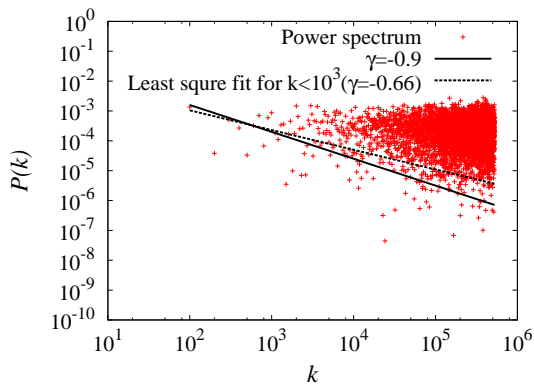}}
\caption{The power spectrum of the sequence generated by FFM with threshold.
The solid line corresponds $\gamma=0.9$.  The broken line
corresponds the least square fitting for data points for $k<10^3$.
The exponent obtained by the fitting is $\gamma=0.66$.
Note that the number of data points is reduced.}
\label{fft.eps}
\end{center}
\end{figure}

The effects of short-range randomness become obvious in the power
spectrum.  Figure \ref{fft.eps} shows the power spectrum of the
sequence generated by FFM with threshold.  The spectrum is almost
flat.  The short-range random noises dominate the spectrum and prevent
us to detect the long-range correlation.

The exponent is obtained as $\gamma \sim 0.02$, if you apply the least
square method for fitting all data points for the spectrum.  The
exponent obtained as $\gamma\sim0.66<0.9$ is smaller than the expected
value, by applying the fitting for long-range data points limited for
$k<k_c$. The power-spectrum analysis, namely, is strongly affected by
short-range noises.  It seems to be difficult to detect long-range
correlations by analyzing the power spectrum.

\begin{figure}[ht]
\begin{center}
\resizebox{0.45\textwidth}{!}{\includegraphics{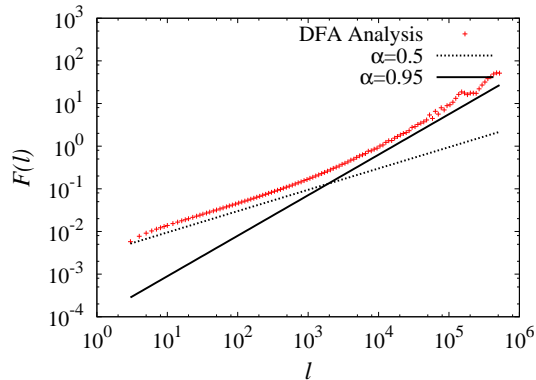}}
\caption{The result of the DFA analysis of the sequence 
generated by FFM with threshold.}
\label{dfa.eps}
\end{center}
\end{figure}

The DFA analysis is more useful in this case than power-spectrum
analyses.  Figure \ref{dfa.eps} shows the result of the DFA analysis.
It shows crossover of two regions.  The exponent is $\alpha\sim0.5$
for the shorter region. It is the value for random sequences.  And for
the longer region it is $\alpha=(\gamma+1)/2=0.95$, which corresponds
to the expected correlation.  The crossover point of these two regions
locates at the threshold $k_c$.  The DFA analysis, namely, detects
the existence of long-range correlation and gives the threshold $k_c$,
above which random noises dominate.

\section{Coarse-grained  Sequence}

Real observed data, in general, will contain various types of
irregularity in short-range area.  The simplest practical way to
eliminate such irregularity is to sum data over some short length.  We
examine that this simple method preserves the long-range correlation
in the original sequence as you expect.

\begin{figure}[ht]
\begin{center}
\resizebox{0.45\textwidth}{!}{\includegraphics{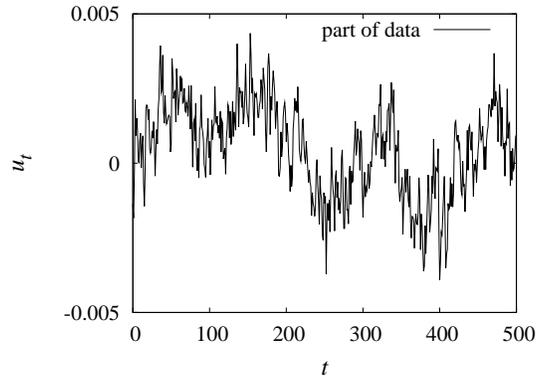}}
\caption{A part of the coarse-grained sequence.
Note that the unit of the $t$-axis is $10^3$ times larger than in 
Fig.\ref{dataWithoutCutoff.eps}.}
\label{dataAveraged.eps}
\end{center}
\end{figure}

The sequence generated in the previous section contains short-range
random noises with long-range correlations.  The DFA analysis gives
the threshold $k_c$, at which short-range noises dominate.  Summing
data over segments of length $T/k_c$ will eliminate those noises.
Figure \ref{dataAveraged.eps} shows the coarse-grained data obtained
by summation up to $T/k_c$.  It shows the existence of long-range
correlations.

\begin{figure}[ht]
\begin{center}
\resizebox{0.45\textwidth}{!}{\includegraphics{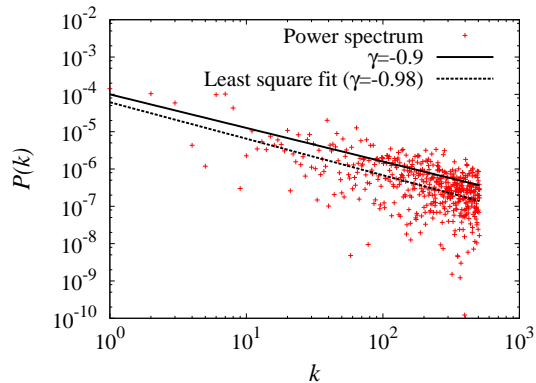}}
\caption{The power spectrum of the coarse-grained sequence.
The solid line corresponds $\gamma=0.9$.  The broken line
corresponds the least square fitting for all data points.
The exponent obtained by the fitting is $\gamma=0.98$.}
\label{fftAveraged.eps}
\end{center}
\end{figure}

Figure \ref{fftAveraged.eps} shows the power spectrum of the
coarse-grained sequence.  Fitting all data by the least square method
gives the exponent $\gamma=-0.98$, which is slightly different from
the imposed one $\gamma=-0.9$.

\begin{figure}[ht]
\begin{center}
\resizebox{0.45\textwidth}{!}{\includegraphics{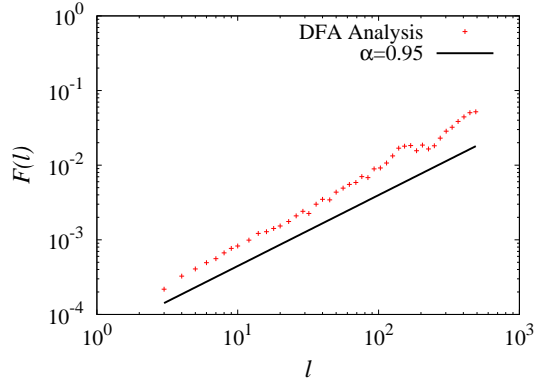}}
\caption{The result of the DFA analysis of the coarse-grained sequence.}
\label{dfaAvereged.eps}
\end{center}
\end{figure}

The DFA analysis for the coarse-grained sequence gives the exponent
$\alpha=0.95$ as shown in Fig.~\ref{dfaAvereged.eps}.  The power-law
correlation with $\gamma = 0.9$ installed into the sequence is
recovered and is detected by the DFA method.

\section{Summary and Discussion}

Power-law correlations contained in temporal data of various dynamical
systems have attracted research interests in various research fields.
Increment of the amount of data sometimes means improvement of
temporal resolution.  Improvement of temporal resolution of data
sometimes introduces asynchronous irregularity into data.  Those
short-range noises may prevent us from analyzing long-term
correlations.

Short-range noises is shown to affect strongly the power-spectrum
analysis.  It is very difficult to detect long-range correlation, if
the data contain such short-range noises.

The detrended fluctuation analysis (DFA) can detect both the
short-range noises and the long-range correlations.  There two ranges
intersect at a threshold.  The DFA also gives the threshold dividing
these two ranges.

Finally we discuss the practicality of this work.  A e-mail service is
one of the most popular services in the Internet.  Users send e-mail
messages to e-mail servers of their own organization or those operated
by Internet service providers. E-mail servers record e-mail sending
requests usually every second.  Every record contains the sender and
receiver addresses and the message size.  Namely the amount of sent
messages is recorded every second.

\begin{figure}[ht]
\begin{center}
\resizebox{0.45\textwidth}{!}{\includegraphics{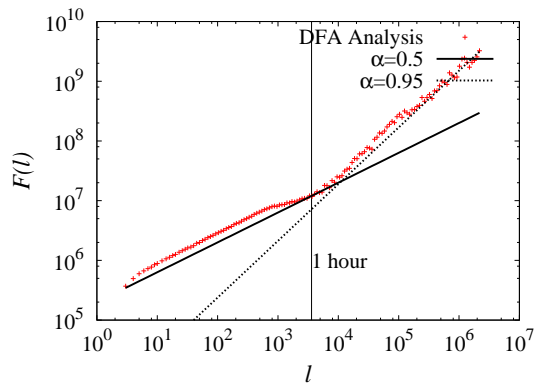}}
\caption{The result of the DFA for e-mail messages at a e-mail server.
In the shorter range than one hour, random noises dominate the data.
The power-law correlation can be found for longer range than one
hour.}
\label{dfa-emailFlow.eps}
\end{center}
\end{figure}

A user will send his message to another user.  The receiver will
respond the message by quoting the received message after some delay.
Therefore the sequence of the amount of e-mail messages will contain
long-range correlations.

Figure \ref{dfa-emailFlow.eps} shows the result of the DFA analysis
for the amount of e-mail messages at a e-mail server.  In the shorter
range than one hour, the exponent is $\alpha\sim 0.5$, which
corresponds to one for random noises.  In the longer range than one
hour, the long range correlation with $\alpha\sim 0.95$ can be found.

The number of the data points of this observation is of order $10^7$.
The short range to the order $10^3$ is dominated random noises.
Namely the study in this paper with artificial sequences will be
applicable to real observed data.  The detail analysis of the cases for
e-mail messages will be discussed elsewhere.

\begin{acknowledgments}
The author would like to thank Y.~Hieida and Y.~Matsubara for their
stimulating discussion.  A part of this work is financially supported
by a Grant-in-Aid for Scientific Research, No.18500215 and No.20360045,
from the Ministry of Education, Culture, Sports, Science and
Technology, Japan.
\end{acknowledgments}

\bibliography{tadaki}

\end{document}